\documentstyle[aps,prl]{revtex}

\def\bv{{\bf v}}
\def\b\sigma{\mbox{\boldmath $\sigma$\unboldmath}}
\def\bu{{\bf u}}

\def\bw{{\bf w}}
\newcommand{\be}{\begin{equation}}
\newcommand{\ee}{\end{equation}}
\def\({\left(}
\def\){\right)}

\begin{document}

\title{Thermalization of a particle by dissipative collisions}

\author{Philippe A. Martin}

\address{Institut de Physique Th\'eorique, Ecole Polytechnique F\'ed\'erale
de Lausanne, CH-1015, Lausanne, Switzerland}
 
\author{Jaros\l aw Piasecki}

\address{Institute of Theoretical Physics, University of Warsaw, Ho\.za 69, 
00 681 Warsaw, Poland}

\date{\today}

\maketitle

\begin{abstract}
One considers the motion of a test particle in an homogeneous fluid in 
equilibrium at temperature $T$, undergoing dissipative collisions with 
the fluid particles. It is shown that the corresponding linear Boltzmann 
equation still posseses a stationary Maxwellian velocity distribution, 
with an effective temperature smaller than $T$.  This effective 
temperature is explicitly given in terms of the restitution parameter 
and the masses.

\end{abstract}

\vspace{5mm}

\vspace{1cm}
The search for stationary states of granular matter has recently been 
a subject of interest for both experimental and theoretical reasons [1,2]. 
Granular matter can be modelized by spherical particles that partially  
dissipate their kinetic energy at collisions. If $(\bu, \bv)$ and 
$(\bar{\bu}, \bar{\bv})$ denote the velocities of two spheres of
mass $m$ and $M$ before and after the collision, they are related by
\begin{eqnarray}
m\bar{\bu}+M\bar\bv &=&m\bu+M\bv
\label{1}\\ 
\hat{\b\sigma}\cdot (\bar{\bv}-\bar{\bu})&=&-\alpha\hat{\b\sigma}
\cdot (\bv-\bu), \;\;\;0<\alpha\leq 1
\label{2}\\
\hat{\b\sigma}^{\bot}\cdot (\bar{\bv}-\bar{\bu})&=&\hat{\b\sigma}^{\bot}
\cdot (\bv-\bu)\nonumber
\end{eqnarray}
where $\hat{\b\sigma}$ is a unit vector normal to the surface of 
the spheres at the point of impact, and $\hat{\b\sigma}^{\bot}$ points 
in the tangent direction: $\hat{\b\sigma}^{\bot}\cdot\hat{\b\sigma}=0$.  
The first relation is the conservation of the center of mass momentum, 
whereas the second one says that the normal component of the relative 
velocity reverses its direction with a  magnitude  reduced by
the factor $\alpha$, the so called restitution parameter
(the tangent component remains unchanged). Solving (\ref{1}),
(\ref{2}) for $\bar{\bu}$ and  $\bar{\bv}$ gives \begin{eqnarray}
\bar{\bu}=\bu+(1-\mu)(1+\alpha)(\hat{\b\sigma}\cdot(\bv-\bu))\hat{\b\sigma},
\;\;\;\; 
\bar{\bv}=\bv-\mu(1+\alpha)(\hat{\b\sigma}\cdot(\bv-\bu))\hat{\b\sigma}
\label{3}
\end{eqnarray}
with $\mu=m/(m+M)$. The inverse relation is obtained by exchanging the 
roles of initial and final velocities and $\alpha$ into $\alpha^{-1}$ 
in (\ref{3}). 

We consider now the motion of a single particle of mass $M$ (the test 
particle) in a fluid of particles of mass $m$ in thermal equilibrium at 
temperature $T$ by means of the linearized Boltzmann equation. All particles 
are spheres of diameter $d$ and the test particle undergoes inelastic 
collisions characterized by the restitution parameter $\alpha$
with the particles of the host fluid. Then, in the homogeneous situation, 
the probability density $f(\bv,t)$ for finding the test
particle at time $t$ with velocity $\bv$  obeys the kinetic equation
$$
\frac{\partial}{\partial t}f(\bv,t)=\rho d^{2}\int d\bu\int d\hat{\b\sigma}
(\hat{\b\sigma}\cdot (\bv-\bu))\theta(\hat{\b\sigma}\cdot (\bv-\bu))
\times\nonumber\\
$$
$$
\left[\alpha^{-2}f\(\bv-\mu\(1+\alpha^{-1}\)(\hat{\b\sigma}
\cdot (\bv-\bu))\hat{\b\sigma},t\)
\phi_{T}\(\bu+(1-\mu)\(1+\alpha^{-1}\)(\hat{\b\sigma}\cdot 
(\bv-\bu))\hat{\b\sigma}\)\right.
$$
\be
\;\;\;\;\;\;\;\;\;\;\;\;\;\;\;\;\;\;\;\;\;\;\;\;\;\;\;\;\;
\;\;\;\;\;\;\;\;\;\;\;\;\;\;\;\;\;\;\;\;\;\;\;\;
\left.-f(\bv,t)\phi_{T}(\bu)\right]
\label{4}
\ee
Here $\rho \phi_{T}(\bu)$ is the  equilibrium state the fluid having uniform 
density $\rho$ and Mawellian velocity distribution
\be
\phi_{T}(\bu)=\(\frac{\beta m}{2\pi}\)^{3/2}\exp\(-\beta\frac{mu^{2}}{2}\),
\;\;u=|\bu|,\;\;\beta=\frac{1}{k_{B}T}
\label{5}
\ee
In (\ref{4}), $\bu$ is the velocity of a particle in the fluid, $\bv$ 
is the velocity of the test particle and the $\hat{\b\sigma}$-integral runs 
on the unit sphere. The velocity arguments of $f$ and $\phi_{T}$ in the gain 
part of the collision term in (\ref{4}) are precisely those that the particles 
must have before the collision to  produce post-collisional velocities $\bu$ 
and $\bv$, according to (\ref{3}). The term$(\hat{\b\sigma}\cdot (\bv-\bu))
\theta(\hat{\b\sigma}\cdot (\bv-\bu))$ (with $\theta(y)$ the Haeviside function) 
reflects the fact that the collision frequency is proportional to the velocity 
of approach of the colliding pair (free motion between encounters). 
The factor $\alpha^{-2}$ compensates for the contractive transformation (\ref{2}) 
in velocity space and ensures that the normalization of $f$ is properly conserved 
in the course of the time.

The point of this letter is to show that, despite of its dissipative 
collisions with the particles of the bath, the distribution of test particle 
still posseses a stationary {\em Maxwellian} velocity distribution, 
with a reduced effective temperature $T^{{\rm eff}}$ that will
be explicitly given in terms of the restitution parameter and the masses. 

We look for a stationary state setting simply $f(\bv,t)=f(\bv)$ and 
equating the r.h.s. of (\ref{4}) to zero. With the change of variable 
$\bw=\bv-\bu$, $f$ has to satisfy
$$
0=\int d\hat{\b\sigma}\int d\bw
(\hat{\b\sigma}\cdot \bw)\theta(\hat{\b\sigma}\cdot \bw)\times\nonumber\\
$$
$$
\left[\alpha^{-2}f\(\bv-\mu\(1+\alpha^{-1}\)(\hat{\b\sigma}\cdot \bw)
\hat{\b\sigma}\)\phi_{T}\(\bv-\bw+(1-\mu)\(1+\alpha^{-1}\)(\hat{\b\sigma}
\cdot\bw)\hat{\b\sigma})\)\right.
$$
\be
\;\;\;\;\;\;\;\;\;\;\;\;\;\;\;\;\;\;\;\;\;\;\;\;\;\;\;\;\;\;\;\;\;\;\;\;\;\;\;
\left.-f(\bv)\phi_{T}(\bv-\bw)\right]
\label{6}
\ee
Let us perform the $\bw-$integration at fixed $\hat{\b\sigma}$ in a frame
\be
\bw=w_{1}\hat{\b\sigma}_{1}+w_{2}\hat{\b\sigma}_{2}+w_{3}\hat{\b\sigma}_{3}
\label{7}
\ee
where $\{\hat{\b\sigma}_{1}=\hat{\b\sigma},\;\hat{\b\sigma}_{2},\;
\hat{\b\sigma}_{3}\}$ is an orthonormal system of unit vectors with 
$\hat{\b\sigma}_{2},\;\hat{\b\sigma}_{3}$ in the plane 
perpendicular to $\hat{\b\sigma}$. Note that the argument of $f$ does 
not depend on the variables $w_{2}$ and $w_{3}$, so the corresponding 
integrals can be carried out directely on $\phi_{T}$. 
In terms of the variables (\ref{7}) we have
\begin{eqnarray}
& &\phi_{T}\(\bv-\bw+(1-\mu)\(1+\alpha^{-1}\)(\hat{\b\sigma}\cdot\bw)
\hat{\b\sigma})\)\nonumber\\
&=&\(\frac{\beta m}{2\pi}\)^{3/2}\exp\left\{-\frac{\beta m}{2}
\(\hat{\b\sigma}\cdot\bv+\(\alpha^{-1}
-\mu\(1+\alpha^{-1}\)\)w_{1}\)^{2}\right\}\nonumber\\
&\times&\exp\(-\frac{\beta m}{2}(\hat{\b\sigma}_{2}\cdot\bv-w_{2})^{2}\)
\exp\(-\frac{\beta m}{2}(\hat{\b\sigma}_{3}\cdot\bv-w_{3})^{2}\) 
\label{8}
\end{eqnarray}
Hence the integrations on $w_{2}$ and $w_{3}$ can be readily performed 
here (as well in $\phi_{T}(\bv-\bw)$) and the stationary equation takes 
eventually the form
$$
0=\int d\hat{\b\sigma}\int_{0}^{\infty}dy y\left[f(\bv-\eta y \hat{\b\sigma})
\exp\(-\frac{\beta m}{2}
(\hat{\b\sigma}\cdot\bv+(1-\eta)y)^{2}\)\right.
$$
\be
\;\;\;\;\;\;\;\;\;\;\;\;\;\;\;\;\;\;\;\;\;\;\;\;
\left.-f(\bv)\exp\(-\frac{\beta m}{2}(\hat{\b\sigma}\cdot\bv-y)^{2}\) \right]
\label{9}
\ee
To obtain (\ref{9}) we also put $w_{1}=\alpha y$ in the first part of 
the integrand in (\ref{6}) (gain term) and we defined the new parameter 
$\eta=\mu(\alpha+1)$.

Now, let us set
$$
f(\bv)=\exp\(-\beta M v^{2}/{2\gamma}\), \; v=|\bv| 
$$
omitting the normalization factor. One notices then that
the following equality holds
\be
M\gamma^{-1}|\bv-\eta y\hat{\b\sigma}|^{2}+m(\hat{\b\sigma}\cdot
\bv+(1-\eta)y)^{2}=M\gamma^{-1}v^{2}+m(\hat{\b\sigma}\cdot\bv-y)^{2}
\label{10}
\ee
provided that $\gamma^{-1}$ has the value 
(independent of $\bv,\;\hat{\b\sigma}$ and $y$)
\be
\gamma^{-1}=\frac{m(2-\eta)}{M\eta}
\label{11}
\ee
The relation (\ref{10}) implies vanishing of the integrand in (\ref{9}). 
Introducing the effective temperature $T^{{\rm eff}}$ by 
\be
T^{{\rm eff}}=\gamma T
\label{12}
\ee
we conclude that the Maxwellian $f(\bv)=\phi_{T^{{\rm eff}}}(\bv)$ is 
a stationary distribution for our test particle.
The explicit formula for the factor $\gamma$ reducing the temperature 
in terms of the restitution parameter and the masses reads 
\be
\gamma=\frac{\alpha+1}{2+(1-\alpha)m/{M}}\leq 1
\label{13}
\ee
Obviously $\gamma=1$ when $\alpha=1$, as it should. For a fixed mass ratio 
$m/M$, $\gamma$ decreases as $\alpha$ is varied from $1$ to $0$. 
Thus it takes its smallest possible value $\gamma=(2+m/M)^{-1}< 1/2$ 
at $\alpha=0$  when the collisions are fully dissipative. 

This result deserves several comments. Usually the dependence of 
the stationary state on the kinematical variables is determined 
by the conservation laws. The stationary Maxwellian velocity
distribution for the hard sphere Boltzmann equation follows from 
the conservation of the total kinetic energy
$\frac{1}{2}mu^{2}+\frac{1}{2}Mv^{2}$  at collisions. It is therefore 
quite remarquable (and not obvious beforehand) that the
distribution remains Gaussian when collisions dissipate the kinetic energy. 
Notice that the integrand in (\ref{9}) vanishes pointwise, that is, 
$\phi_{T^{{\rm eff}}}(\bv)$ is also stationnary in situations
where the collision frequency, proportional to  
$(\hat{\b\sigma}\cdot \bw)\theta(\hat{\b\sigma}\cdot \bw)$, is replaced 
by a more general positive function of  $\hat{\b\sigma}\cdot \bw$.
A particular case is the so called Maxwell gas for which the collision 
frequency is independent of the velocity. Moreover, the result holds 
in any space dimension $d=1,2,3$ with the same formulae
(\ref{12}), (\ref{13}) for the effective temperature.

These findings must be contrasted to other physical situations 
where the stationary distribution is definitely not 
Maxwellian. One example is the Lorentz model with dissipative 
collisions: a particle, acted upon an uniform external field,
undergoes inelastic collisions with randomly distributed 
static (infinitely heavy) scatterers. Here the dissipation
(however weak it may be) sufficies to balance the energy 
flow from the external field and to guarantee the existence of
a stationary state. In the weakly dissipative regime, the stationary 
velocity distribution is found to behave as $\sim\exp(-(1-\alpha)v^{4}/2)$ 
[3,4]. The physics of the Lorentz gas is not the same
since there is no thermalization mechanism available for the moving 
particle. In [5], one considers a one-dimensional system of
infinitely many inelastic particles in contact with a thermal 
bath, modelized by a Fokker-Plank equation in the limit
of weak dissipation. In this case, the stationary velocity distribution 
behaves as $\sim\exp(-Cv^{3})$, but here dissipation occurs internally 
within the many particle system, and not in relation with 
its interaction with the thermal bath.

Further investigations of interest are the study of 
the approach to the steady state in the course of time
and the possible generalization of our result to a 
system of several interacting test particles.

\vspace{5mm}
{\bf Acknowledgements}

\vspace{2mm}
J.Piasecki acknowledges the hospitality at the Institute of 
Theoretical Physics of the Ecole Polytechnique F\'ed\'erale de 
Lausanne (Switzerland), and financial support by the KBN 
(Committee for scientific Research, Poland), grant 2 P03B 03512.

\end{document}